\documentclass[12pt]{iopart}

\usepackage{graphicx}
\def\prl{Phys. Rev. Lett.}
\def\prd{Phys. Rev. D}
\def\cqg{Class. Quantum Grav.}

\begin{document}

\title[The stationary 1+log slicing]{Equilibrium initial data for moving puncture simulations: The stationary 1+log slicing}

\author{T. W. Baumgarte$^{1,2}$, Z. B. Etienne$^2$, Y. T. Liu$^2$, K. Matera$^1$, N. \'{O} Murchadha$^3$, S. L. Shapiro$^{2,4,5}$ and K. Taniguchi$^2$}

\address{${}^1$ Department of Physics and Astronomy, Bowdoin College,  
Brunswick, ME 04011, USA}

\address{${}^2$ Department of Physics, University of Illinois at
  Urbana-Champaign, Urbana, Illinois 61801, USA}

\address{${}^3$ Physics Department, University College, Cork, Ireland}

\address{${}^4$ Department of Astronomy, University
  of Illinois at Urbana-Champaign, Urbana, Illinois 61801, USA}
  
\address{${}^5$ NCSA, University
  of Illinois at Urbana-Champaign, Urbana, Illinois 61801, USA}
  
\begin{abstract}
We discuss a ``stationary 1+log" slicing condition for the construction of solutions to Einstein's constraint equations.  For stationary spacetimes, these initial data give a stationary foliation when evolved with ``moving puncture" gauge conditions that are often used in black hole evolutions.  The resulting slicing is time-independent and agrees with the slicing generated by being dragged along a time-like Killing vector of the spacetime.   When these initial data are evolved with moving puncture gauge conditions, numerical errors arising from coordinate evolution should be minimized.  While these properties appear very promising, suggesting that this slicing condition should be an attractive alternative to, for example, maximal slicing, we demonstrate in this paper that solutions can be constructed only for a small class of problems.  For binary black hole initial data, in particular, it is often assumed that there exists an approximate helical Killing vector that generates the binary's orbit.  We show that 1+log slices that are stationary with respect to such a helical Killing vector cannot be asymptotically flat, unless the spacetime possesses an additional axial Killing vector.
\end{abstract}

\section{Introduction}

Numerical simulations of binary black hole systems have recently experienced a dramatic breakthrough (see \cite{Pre05b,BakCCKM06a,CamLMZ06} as well as numerous follow-up publications).   Most of these simulations now adopt some variation of the BSSN formulation \cite{ShiN95,BauS99} together with the moving puncture method to handle the black hole singularities.  

The original puncture method, which factors out the singular behavior of black hole spacetimes analytically, was originally developed in the context of initial data \cite{BeiO94,BeiO96,BraB97}.  
When used for dynamical simulations (e.g.~\cite{Bru99,AlcBBLNST01,BruTJ04}), this method did not lead to long-term stable evolutions.    The breakthrough in the recent puncture simulations is based on a ``moving puncture" approach, in which the singular terms are no longer factored out, and in which the punctures are allowed to propagate through the numerical grid.  With a set of empirically determined coordinate conditions, and as long as no grid point ever encounters a singular term directly, this prescription then leads to remarkably stable evolutions.   

It is quite surprising that the presence of singularities does not spoil the numerical calculation.   This issue has been clarified by Hannam et al.~\cite{HanHPBO06,HanHOBGS06,HanHOBO08} (see also \cite{Bro07}), who analyzed the geometry of moving puncture evolutions for single Schwarzschild solutions.  Typically, such evolutions start out with a slice of constant Schwarzschild time expressed in isotropic coordinates as initial data.  These coordinates do not penetrate the black hole interior, and instead cover two copies of the black hole exterior, corresponding to two sheets of asymptotically flat ``universes'', connected by an Einstein-Rosen bridge at the black hole horizon.  Given their appearance in an embedding diagram, these data are often called ``wormhole" data.  The singularity at isotropic radius $r=0$, where the conformal factor $\psi = 1 + M/(2r)$ diverges, corresponds to the asymptotically flat end of the ``other'' universe, and is therefore a coordinate singularity only.  When evolved with the ``1+log" slicing condition for the lapse \cite{BonMSS95} and a ``$\bar \Gamma$-freezing" gauge condition for the shift \cite{AlcBDKPST03}, the solution passes through a short phase of ``dynamical" evolution, but then settles down into a new equilibrium solution.  An embedding diagram of this new equilibrium solution is shown in Fig.~2 of \cite{HanHOBO08} and suggests the name ``trumpet" data.  The solution terminates at a non-zero areal radius, and therefore does not encounter the spacetime singularity at the center of the black hole.   The singularity at the isotropic radius $r=0$ is therefore again a coordinate singularity only. This new equilibrium is a stationary slice of the ``1 + log'' slicing condition relative to the standard timelike Killing vector of the Schwarzschild solution.

At least in the case of Schwarzschild, the brief ``dynamical" phase of the evolution refers to a coordinate evolution only, since the spacetime itself is clearly static.   Even though the spacetime is static, it is sliced, in this case, by  a nonstationary foliation.   The resulting nontrivial evolution of the coordinates leads to a time-dependence of the metric coefficients, which may well introduce or enhance numerical errors in dynamical simulations.    This observation suggests that it would be advantageous to start the numerical evolution not with slices of constant Schwarzschild time as initial data, but instead with a slicing that corresponds to the late-time equilibrium solution.  Stated differently, we would like to identify a ``stationary" slicing of the spacetime that satisfies the 1+log slicing condition used in moving puncture simulations.  All quantities should then remain constant during a time evolution (up to numerical truncation error), potentially reducing numerical errors and noise.  

For binary black hole evolutions, moving puncture simulations have typically used maximally sliced puncture initial data (see, e.g., \cite{BraB97,Bau00}).  In the limit of infinite binary separation, these data again approach slices of constant Schwarzschild time for two individual static black holes (see \cite{DenBP06} for a perturbative analysis of this limit).  Presumably, these data contain several sources of error, including some nonzero eccentricity (see \cite{PfeBKLLS07,HusHGSB08}) and the absence of the correct gravitational wave pattern originating from the prior binary inspiral.  Another potential source of error arises from the fact that the individual black holes are not in equilibrium when evolved with the 1+log and $\tilde \Gamma$-freezing conditions, even at infinite binary separation.  As described above for single black holes, a dynamical evolution will lead to the individual black holes evolving to a new quasiequilibrium coordinate state.  This ``coordinate-dynamical" process may again introduce numerical error, and may contribute to the ``spurious gravitational radiation" that is often observed at early times in binary black hole simulations.

It would again be desirable to construct initial data that describe black holes in a coordinate system that is already in equilibrium with respect to 1+log and $\tilde \Gamma$-freezing coordinates used in moving puncture simulations.  Obviously, stationary coordinate systems can only exist for stationary spacetimes.  While inspiralling binary systems are not stationary, they possess an approximate helical Killing vector, and it would be desirable to find a ``stationary 1+log" slicing with respect to this Killing vector.  As we will discuss below, a corresponding slicing condition can be formulated quite naturally in the context of the conformal thin-sandwich decomposition of Einstein's constraint equations.  Given these considerations, this ``stationary 1+log" slicing condition would appear as an attractive alternative to maximal slicing, which is often used in the construction of binary black hole initial data.  However, we demonstrate in this paper that, in the absence of an axial Killing vector, asymptotically flat solutions to the resulting equations cannot exist.  In particular this rules out solutions describing binary black holes.

 This paper is organized as follows.  In Section \ref{sec:BE} we briefly review the basic equations, including the 1+log slicing condition and the conformal thin-sandwich decomposition of Einstein's constraint equations.  We then solve the latter equations, subject to the 1+log slicing condition, for Schwarzschild black holes in Section \ref{sec:SS}.  In Section \ref{sec:binaries} we then demonstrate why such solutions cannot exist in rotating spacetimes, in the absence of an axial Killing vector.  We stress that this nonexistence result is valid only where the only Killing vector is a helical one.  If the Killing vector is a combination of a rotational Killing vector and a time-translational one, the problem disappears. We briefly summarize in Section \ref{sec:summary}.  In Appendix \ref{sec:SS} we also include some results from dynamical evolution simulations of our Schwarzschild solutions.

\section{Basic equations}
\label{sec:BE}

\subsection{The 3+1 decomposition}

We write the spacetime metric $g_{ab}$ in the 3+1 form
\begin{equation}
g_{ab} dx^a dx^b = - \alpha^2 dt^2 + \gamma_{ij} (dx^i + \beta^i dt) 
(dx^j + \beta^j dt),
\end{equation}
where $\alpha$ is the lapse function, $\beta^i$ the shift vector, and  $\gamma_{ij}$ the spatial metric induced on a spatial slice $\Sigma$ of constant coordinate time $t$.  The normal on this slice is $n^a = \alpha^{-1} (1, -\beta^i)$ (note that in our ''Fortran" convention letters $a, b, c, \ldots$ denote spacetime indices, while letters $i, j, k, \ldots$ denote spatial indices).  We also define the extrinsic curvature as $K_{ij} = - (1/2) {\mathcal L}_{\bf n} \gamma_{ij}$, where ${\mathcal L}_{\bf n}$ denotes the Lie derivative along $n^a$ (we adopt the 
``ADM" convention of Arnowitt, Deser and Misner \cite{ArnDM62} rather than that of \cite{Wal84}).  Expanding this Lie derivative we have
\begin{equation} \label{eq:gamma_dot}
\partial_t \gamma_{ij} = - 2 \alpha K_{ij} + D_i \beta_j + D_j \beta_i,
\end{equation}
where $D_i$ is the covariant derivative associated with $\gamma_{ij}$. 

Einstein's equations can then be decomposed into two sets of equations for the field variables $\gamma_{ij}$ and $K_{ij}$.  One of these sets is the constraint equations, namely the Hamiltonian constraint and the momentum constraint, which constrain the field variables on each spatial slice $\Sigma$.  The second set of equations is the evolution equations, one of which is given by equation (\ref{eq:gamma_dot}), which determines the time evolution of the fields from one spatial slice to the next.  The lapse and the shift, which determine how the coordinates evolve from one spatial slice to the next, appear only in the evolution equations, and have to be chosen independently before the evolution can proceed.

Constructing initial data entails finding solutions to the two constraint equations.  Before this can be done, a particular decomposition of  the constraint equations has to be chosen (see, e.g., \cite{Coo00,BauS03,Gou07b} for reviews).  Particularly popular decompositions are the conformal transverse-traceless decomposition, which provides the framework for the black hole puncture initial data (see \cite{BeiO94,BeiO96,BraB97,Bau00}), and the conformal thin-sandwich decomposition.  As we will see below, the latter is more suitable for our purposes.

\subsection{Stationary 1+log slicing}
\label{sec:1+log}

The 1+log slicing condition, which is often employed in dynamical moving puncture simulations, is usually written as
\begin{equation}  \label{1+log_dyn}
(\partial_t - \beta^i \partial_i) \alpha = - 2 \alpha K
\end{equation}
(see \cite{BonMSS95}).  Here the mean curvature $K$ is the trace of the extrinsic curvature, $K = \gamma^{ij} K_{ij}$.  We point out that the presence of the advective shift term allows us to write the left hand side as a derivative along the normal vector, 
\begin{equation}
n^a \nabla_a \alpha = {\mathcal L}_{\bf n} \alpha = - 2 K.
\end{equation}
This means that this slicing is ``covariant" in the sense that it does not depend on the choice of the shift -- for a given slice, and a given initial lapse, the resulting slicing of the spacetime is unique.


We would now like to construct a stationary slicing that satisfies the slicing condition (\ref{1+log_dyn}).  Clearly we can find a stationary slicing of a spacetime only if the spacetime is stationary, meaning that the latter possesses a Killing vector $\xi^a$.  A stationary slicing of the spacetime is then dragged along by $\xi^a$.  We can now construct a coordinate system by identifying the time coordinate vector with the Killing vector $\xi^a$, i.e.
\begin{equation}
\xi^a = t^a = \alpha_K n^a + \beta_K^a.
\end{equation}
This identification provides what we call the Killing lapse $\alpha_K$ and the Killing shift $\beta^a_K$.    Given a certain initial slice, and evolving this slice with the Killing lapse and shift will render all quantities independent of time.  In the ``1 +log'' slicing, the initial value of the lapse is a free choice, and the foliation condition determines its time evolution.  To construct a stationary slicing, the initial lapse must be the Killing lapse. The slicing lapse must remain equal to the Killing lapse when dragged along the Killing vector, i.e., when one chooses the slicing shift to be equal to the Killing shift. However, whether or not this Killing lapse simultaneously satisfies the 1+log slicing condition (\ref{1+log_dyn}) depends on the choice of our initial slice.

For a concrete example, consider the Schwarzschild spacetime, foliated by slices of constant Schwarzschild time $t$, in isotropic coordinates.   We can then identify the Killing lapse as 
\begin{equation} \label{alpha_K_SS}
\alpha_K = \frac{1-M/(2r)}{1+M/(2r)}
\end{equation}
and the Killing shift as $\beta_K^r = 0$.  Given that the spatial metric is time-independent on these slices we also have $K_{ij} =0 = K$, and we can readily verify that this slicing does satisfy the 1+log slicing condition (\ref{1+log_dyn}).  Note, however, that this $\alpha_K$ is negative for $r < M/2$. We can just as easily construct a counter example.  In Painlev\'e-Gullstrand coordinates we can identify the Killing lapse as $\alpha_K = 1$, the Killing shift as $\beta_K^r = (2M/r)^{1/2}$, and the mean curvature is $K = (3/2) (2M/r^3)^{1/2}$.  Evidently, slices of constant Painlev\'e-Gullstrand times do not satisfy (\ref{1+log_dyn}), nor do slices of constant Kerr-Schild or ingoing Eddington-Finkelstein time.

Having chosen the slicing lapse $\alpha_S$ as equal to the Killing lapse, and the slicing shift equal to the Killing shift, we need $\partial_t \alpha_K = \partial_t \alpha_S = 0$. The condition (\ref{1+log_dyn}) then reduces to
\begin{equation} \label{1+log}
K = \frac{\beta_K^i \partial_i \alpha_K}{2 \alpha_K}.
\end{equation}
We will refer to this condition as ``stationary 1+log slicing".

In the following we will employ the condition (\ref{1+log}) for the construction of initial data.  As we discussed above, the constraint equations alone do not provide conditions for the lapse and shift, so that it is not clear how (\ref{1+log}) could be employed in any meaningful way.  The conformal thin-sandwich decomposition, however, is constructed by imposing conditions on the time evolution of the spatial metric, which automatically introduces the lapse and shift.  This suggests that we should adopt the conformal thin-sandwich decomposition to solve the constraint equations in the context of the stationary 1+log slicing condition.

As an aside -- which will provide a useful analogy later in this paper --  we note that some authors have considered the condition (\ref{1+log_dyn}) without the advective term,
\begin{equation} \label{1+log_NS}
\partial_t \alpha = - 2 \alpha K
\end{equation}
(which is no longer a ``covariant" slicing condition in the sense that we discussed above).  In this case, the Killing lapse associated with a stationary slicing satisfies this condition only if the slices are maximal, i.e.~if the mean curvature vanishes $K=0$.  For Schwarzschild, all maximal slicings can be parametrized by a parameter $C$ (see \cite{EstWCDST73,Rei73,BeiO98}).  The lapse, for example, is given by
\begin{equation} \label{lapse_MS}
\alpha_K = \left( 1 - \frac{2M}{R} + \frac{C^2}{R^4} \right)^{1/2},
\end{equation}
where $R$ is an areal radius (in contrast to the isotropic radius $r$ in (\ref{alpha_K_SS})).  For $C=0$ we recover the slices of constant Schwarzschild time $t$, as in (\ref{alpha_K_SS}).  As discussed by \cite{CooP04}, we could also parametrize this family of slicings by the value of the lapse on the horizon, $\alpha_{\rm AH}$.  For $C=0$, for example, we evidently have $\alpha_{\rm AH} = 0$.

Dynamical simulations of Schwarzschild spacetimes, adopting the condition (\ref{1+log_NS}) and a $\tilde \Gamma$-freezing condition for the shift, settle down not to an arbitrary maximal slice, but instead to that with $C = 3 \sqrt{3} M^2/4$, which has a limiting surface at $R = 3M/2$ (see \cite{HanHOBGS06}).  
To construct initial data that lead to a trivial time-evolution under the condition (\ref{1+log_NS}), we would therefore use this particular maximal slice (see \cite{BauN07}).    We will return to this issue in Section \ref{sec:SS}.

\subsection{The conformal thin-sandwich decomposition}
\label{sec:CTS}

We begin by conformally decomposing the spatial metric as
\begin{equation}\label{metric}
\gamma_{ij} =  \psi^4 \bar \gamma_{ij},
\end{equation}
where $\psi$ is a conformal factor and $\bar \gamma_{ij}$ a conformally related metric.  
We split the extrinsic curvature $K_{ij}$ into its trace $K$ and a traceless part $A_{ij}$ according to
\begin{equation}
K_{ij} =A_{ij} +\frac{1}{3} \gamma_{ij}  K \equiv \psi^{-2} \bar A_{ij} +  
\frac{1}{3} \gamma_{ij} K.
\end{equation}
In vacuum, the Hamiltonian constraint then becomes an equation for the conformal factor
\begin{eqnarray}\label{psiConstraint}
\bar D^2 \psi = \frac{1}{8} \psi \bar{R} +  \frac{1}{12}\psi^5 K^2 - 
\frac{1}{8} \psi^{-7} \bar A_{ij} \bar A^{ij}.
\end{eqnarray}
Here $\bar D_i$ and $\bar{R}$ are the covariant derivative and the  
Ricci scalar associated with the conformally related metric $\bar \gamma_{ij}$, and the covariant Laplace operator is  $\bar D^2 = \bar \gamma^{ij} \bar D_i \bar D_j$.   

In the conformal thin-sandwich decomposition (see \cite{Yor99,PfeY03}, as well as \cite{Coo00,BauS03,Gou07b} for reviews) we use the evolution equation for the spatial metric (\ref{eq:gamma_dot}) to express $\bar A^{ij}$ as
\begin{equation}\label{matrixA}
\bar A^{ij} = \frac{1}{2\bar \alpha} \left( \left( \bar L \beta  
\right)^{ij}-\bar u^{ij} \right).
\end{equation}
Here $\bar \alpha = \psi^{-6} \alpha$ and $\bar{u}^{ij} = \partial_t  
\bar \gamma_{ij}$, and the conformal Killing operator  $\bar{L}$ is defined as
\begin{equation}\label{Lbar}
\left( \bar L \beta \right) ^{ij} \equiv \bar D^i \beta^j + \bar D^j  
\beta^i -\frac{2}{3} \bar \gamma^{ij} \bar D_k \beta^k .
\end{equation}
The momentum constraint then becomes
\begin{equation}\label{betaConstraint}
\left( \bar \Delta_L \beta \right)^i =\left( \bar L \beta \right)^ 
{ij} \bar D_j \ln \left( \bar \alpha \right)+
\bar \alpha \bar D_j \left( \bar \alpha^{-1} \bar u^{ij} \right)  
+\frac{4}{3} \bar \alpha \psi^6 \bar D^i K,
\end{equation}
where $(\bar \Delta_L \beta)^i = D_j (\bar L \beta)^{ij}$ is a vector  
Laplacian.  In the extended conformal thin-sandwich formalism we also combine the trace of the  
evolution equation for $K_{ij}$ with the Hamiltonian  
constraint to find an equation for the lapse $\alpha$,
\begin{equation}\label{alphaConstraint}
\bar{D}^2 \left( \alpha \psi \right) = \alpha \psi \left( \frac{7} 
{8} \psi^{-8} \bar A_{ij} \bar A^{ij} + \frac{5}{12} \psi^4 K^2 +  
\frac{1}{8}\bar{R} \right) 
- \psi^5 \partial_t  
K + \psi^5 \beta^i \bar{D}_i K. 
\end{equation}

The above equations form a set of equations for the lapse $\alpha$,  the shift $\beta^i$ and the conformal factor $\psi$.  Before these equations can be solved, however, we have to make choices for the freely specifiable quantities for the conformal metric and its time derivative, $\bar \gamma_{ij}$ and $\bar u_{ij} =  \partial_t \bar \gamma_{ij}$, as well as for the mean curvature and its time derivative, $K$ and $\partial_t K$.   

To motivate our choices, consider a given stationary spacetime that, by virtue of being stationary, possesses a Killing vector.  Let us choose an arbitrary spacelike slice through this spacetime.  Projecting the spacetime metric $g_{ab}$ onto this slice yields the spatial metric $\gamma_{ij}$ and, given the normal vector $n^a$, we can also compute the slice's extrinsic curvature $K_{ij}$ as well as its trace $K$.  To assemble ``conformal thin sandwich" data on this slice we also have to choose a time direction, which amounts to choosing a lapse and a shift.  Aligning this time direction with the Killing vector, which yields the Killing lapse and shift, leads to a stationary slicing of the our stationary spacetime.  We then have $\partial_t \gamma_{ij} = 0$ and $\partial_t K = 0$.   We also choose a conformal factor, so that we can compute the conformally related metric $\bar \gamma_{ij}$, and furthermore choose the conformal factor to be independent of time.  This guarantees that $\bar u_{ij} =0$.   Clearly, then, our conformal factor $\psi$ as well as the Killing lapse and shift must be solutions to the conformal thin-sandwich equations for the given $\bar \gamma_{ij}$ and $K$, as well as $\bar u_{ij} =0$ and $\partial_t K = 0$.  Reversing the argument, we see that we can find this conformal factor and the Killing lapse and shift 
as solutions to the conformal thin-sandwich equations if we specify the given $\bar \gamma_{ij}$ and $K$, as well as $\bar u_{ij} =0$ and $\partial_t K = 0$ as the freely specifiable quantities.

To construct a stationary slice, or at least an approximately stationary slice through an approximately stationary spacetime, we should therefore choose  $\bar u_{ij} =0$ and $\partial_t K = 0$ together with some $\bar \gamma_{ij}$ and a $K$.  Clearly, these choices cannot guarantee that the resulting slice will be stationary, since the conformal thin-sandwich formalism does not allow us to impose conditions on the first time derivative of the conformal factor,  or the first time derivative of the tracefree part of the extrinsic curvature.  However, the data is in some partial sense stationary because both $\partial_t \bar \gamma_{ij} = 0$ and $\partial_t K = 0$.  Choosing the mean curvature according to
\begin{equation} \label{1+log(2)}
K = \frac{\beta^i \partial_i \alpha}{2 \alpha},
\end{equation}
in analogy to (\ref{1+log}),  further ensures $\partial_t\alpha = 0$ if we were to evolve these data using the 1 + log slicing condition (\ref{1+log_dyn}).   We propose this condition as a ``stationary 1+log" slicing condition.  In the following  we will also assume conformal flatness, $\bar  \gamma_{ij} = \eta_{ij}$, where $\eta_{ij} $ is the flat metric in whatever coordinate system we choose. This conformal flat assumption plays no real role in our result.  

In addition to making choices for the freely specifiable quantities, we also need to impose suitable boundary conditions before the above equations can be solved.  We will assume asymptotic flatness at infinity, which corresponds to 
\numparts
\begin{eqnarray} \label{bc_infinity}
\lim_{r \rightarrow \infty} \psi & = & 1\\
\lim_{r \rightarrow \infty} \alpha & = & 1. 
\end{eqnarray}
We will allow for a rotating coordinate system, so that the outer boundary condition on the shift becomes
 \begin{equation}\label{bc_shift}
 \lim_{r \rightarrow \infty} \beta^i  =  \epsilon^{ijk} \Omega_j x_k.
 \end{equation}
 \endnumparts
We construct black holes by excising the interior of coordinate spheres, and imposing the black hole equilibrium boundary conditions of \cite{CooP04} on the surfaces of these spheres.  These conditions determine the conformal factor and shift on the black holes' horizons, but leave the lapse $\alpha_{\rm AH}$ undetermined.  We discuss our choice for the horizon lapse below.

\section{Schwarzschild black holes}
\label{sec:SS}

Before exploring the slicing condition (\ref{1+log(2)}) for rotating spacetimes in Section \ref{sec:binaries}, we will first examine non-rotating, spherically symmetric Schwarzschild spacetimes in this Section.

To construct initial data, we solve equations (\ref{psiConstraint}), (\ref{betaConstraint}) and (\ref{alphaConstraint}) under the assumption of spherical symmetry with $\bar u_{ij} = 0 = \partial_t K$, adopting the algorithm and numerical code described in \cite{MatBG08}.   The boundary conditions at spatial infinity are given by (\ref{bc_infinity}) with $\Omega_j = 0$ for nonrotating spacetimes.  We also excise the black hole interior at a coordinate location $r_{\rm AH}$.  Following \cite{CooP04} we require the black hole to be in equilibrium, which results in boundary conditions on the conformal factor $\psi$ and the shift $\beta^r$; the lapse $\alpha_{\rm AH}$, however, can be chosen arbitrarily.   

This situation is completely analogous to the maximal slices that we discussed at the end of Section \ref{sec:1+log}.  Had we imposed maximal slicing instead of the stationary 1+log slicing (\ref{1+log(2)}), then each value of $\alpha_{\rm AH}$ would correspond to one particular parameter $C$ in the family (\ref{lapse_MS}) of maximal slices of Schwarzschild.  Since dynamical simulations settle down to the solution with $C = 3 \sqrt{3}M^2/4$, we would want to pick this particular solution by choosing $\alpha_{\rm AH} = 3 \sqrt{3}/16$.

\begin{figure}[t]
\includegraphics[width=3in]{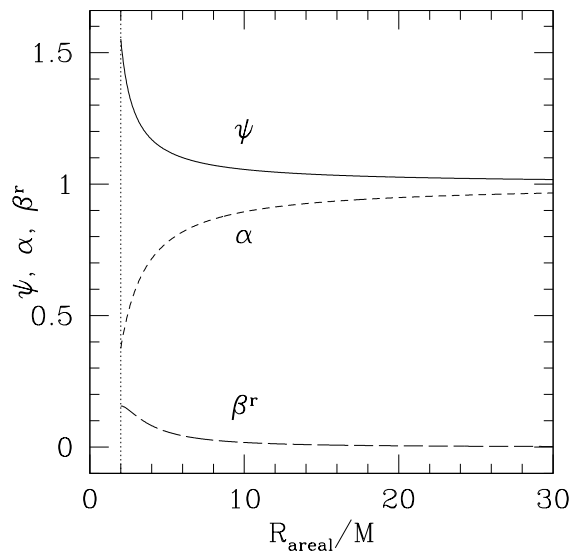}
\caption{Graphs of the conformal factor $\psi$, the lapse $\alpha$, and the shift $\beta^r$ for Schwarzschild in the stationary 1+log slicing (\ref{1+log(2)}).}
\label{Fig1}
\end{figure}

For the stationary slicing (\ref{1+log(2)}) we see that we can again produce a family of solutions, parametrized by the horizon lapse $\alpha_{\rm AH}$.  We again want to pick that member of this family to which the dynamical simulations settle down.  This particular solution has been identified in \cite{HanHPBO06}, and we use this results to identify the corresponding value of $\alpha_{\rm AH}$.
By inspection of their equation (9) we see that the lapse and the areal radius must simultaneously 
satisfy the equations
\numparts
\begin{eqnarray}
3M/R - 2 + 2 \alpha^2 & = & 0\\
2M/R - 1 + \alpha^2 - 2 \alpha & = & 0
\end{eqnarray}
\endnumparts
at a certain critical radius for the solution to remain regular.  Here $M$ is the total mass of the spacetime, and, as before, $R$ is the areal Schwarzschild radius.  Solving these two equations yields $\alpha_c = \sqrt{10} - 3$ and $R_c = M/(4(\sqrt{10} - 3))$.  In spherical symmetry, the slicing condition (\ref{1+log(2)}) has a first integral
\begin{equation} \label{first_int}
\alpha^2 = 1 - \frac{2M}{R} + \frac{C^2}{R^4}  e^\alpha
\end{equation}
(see \cite{HanHOBO08}).  Here $C$ is a constant of integration, which plays the same role as the constant $C$ in (\ref{lapse_MS}).  In fact, this solution differs from (\ref{lapse_MS}) only in the exponential term.  We now insert the above values of $\alpha_c$ and $R_c$ into (\ref{first_int}) to find the value of $C$ for the solution of interest 
\begin{equation}
C = \frac{\sqrt{2}}{16(\sqrt{10} - 3)^{3/2}} e^{(3 - \sqrt{10})/2} M^2 \approx 1.24672 M^2.
\end{equation}
Finally, we use this value together with $R=2M$ in (\ref{first_int}) to obtain an equation for our value of  the horizon lapse.  Solving this equation iteratively yields
\begin{equation}
\alpha_{\rm AH} \approx 0.376179.
\end{equation}
We adopt this value as a Dirichlet boundary condition for our differential equation.  The solution of these equations is now completely determined, and we show numerical results for the conformal factor $\psi$, the lapse $\alpha$ and the shift $\beta^r$ in Fig.~\ref{Fig1}.   An alternative method for constructing these ``trumpet" data is discussed in \cite{HanHOBO08}, who also provide an embedding diagram in their Fig.~2.

In Appendix \ref{sec:SS_evolve} we present some results from dynamical simulations of our Schwarzschild initial data and compare with evolutions of ``puncture" data.

\section{Rotating spacetimes}
\label{sec:binaries}

After exploring spherically symmetric spacetimes under the stationary 1+log condition (\ref{1+log(2)}), we next tried to construct solutions for binary black holes.  Unfortunately, however, we were unable to do so.  We have reason to believe that the problem is not of numerical nature -- using maximal slicing $K=0$ instead of (\ref{1+log(2)}), for example, our code reproduces the results of \cite{CooP04,CauCGP06} without problem -- and instead believe that for rotating, non-axisymmetric spacetimes the condition (\ref{1+log(2)}) does not allow any asymptotically flat solutions.  We support this hypothesis with the following arguments.

\subsection{General considerations}

A standard approach to constructing `quasi-stationary' initial data to the binary black hole problem goes as follows: One selects the independent data by specifying a background (conformal) metric and setting  $\bar u_{ij} = \partial_t \bar \gamma_{ij} = 0$ and $\partial_t K = 0$.  Because of the way the black holes orbit each other, one expects the best approximation to a Killing vector to be helical.   Presumably, then, the time derivatives of $\bar \gamma_{ij}$ and $K$ are close to zero in a corotating coordinate system, which motivates our boundary conditions (\ref{bc_infinity}).  The orbital angular velocity $\Omega_j$ then appears in the equations as a free parameter that can be adjusted until a relativistic virial theorem is satisfied -- namely the equality of the Komar mass  with the ADM mass (see \cite{GouB94}) -- indicating that the binary is in an approximately circular orbit.  Most applications have adopted maximal slicing $K = 0$; here we consider the ``stationary 1+log" slicing condition (\ref{1+log(2)}) instead.

\subsection{Spherically symmetry spacetimes}
\label{sec:bin-SS}

Before proceeding, it is instructive to return to the spherically symmetric spacetimes of Section \ref{sec:SS}.

As above we will assume that the conformal factor falls off to one at infinity according to
\begin{equation} \label{psi_falloff}
\psi = 1 + \frac{M}{2r} + \cdots,
\end{equation}
where $M$ is the spacetime's ADM mass,
and will analyze the asymptotic behavior of the lapse $\alpha$ and the shift $\beta^i$.  Specifically, we 
will inspect the lapse equation (\ref{alphaConstraint}).   It is cleaner to assume that we have found a solution and use the conformal factor to transform to the solution frame. In this case the lapse equation becomes
\begin{equation} \label{eq:lapse}
D^2 \alpha = \alpha K^{ij}K_{ij} +  \beta^i \partial_i K.
\end{equation}
The solution of this equation can be written as a sum of a complementary solution $\alpha_{\rm comp}$ that satisfies the associated homogeneous equation, and a particular solution $\alpha_{\rm part}$ that ``responds" to the source term,
\begin{equation} \label{alpha_ansatz}
\alpha = \underbrace{1 + \frac{M_\alpha}{r} + \frac{Q_\alpha}{r^3} + \cdots}_{\alpha_{\rm comp}}
	+ \underbrace{\frac{S_\alpha}{r^n} + \cdots}_{\alpha_{\rm part}}.
\end{equation}
Here $M_{\alpha}$ is a constant monopole moment, we assume that the dipole moment vanishes (which a suitable coordinate transformation will always accomplish), and we note that the quadrupole moment $Q_{\alpha}$ has the standard flat-space harmonic function.  We will assume that $\alpha$ falls off to one at infinity, i.e.~we will assume $n > 0$.  
Inserting the ansatz (\ref{alpha_ansatz}) into (\ref{eq:lapse}) we see that the left hand side scales according to
\begin{equation} \label{lhs}
D^2 \alpha \sim \frac{S_\alpha}{r^{n+2}},
\end{equation}
where the symbol $\sim$ indicates that asymptotically the leading-order term of the left hand side is proportional to the right hand side.

With the assumption $n > 0$ in (\ref{alpha_ansatz}), the condition (\ref{1+log(2)}) reduces to
\begin{equation} \label{K}
K \sim \beta^i \partial_i \alpha,
\end{equation}
Finally we assume that we can write the shift in the form 
\begin{equation} \label{beta_ansatz}
\beta^i = \epsilon^{ijk} \Omega_j x_k + \frac{B^i}{r^m} + \cdots
\end{equation}
in Cartesian coordinates.  Here the first term on the right-hand side is purely angular and accommodates the transformation from an inertial coordinate system into a rotating one.

\begin{figure}[t]
\includegraphics[width=3in]{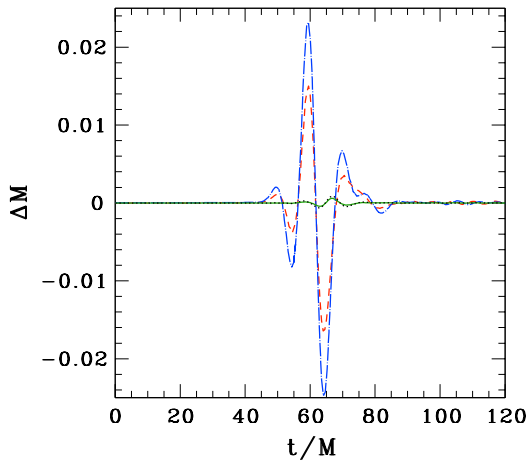}
\caption{The asymptotic behavior of the shift $\beta$ and the mean curvature $K$ in spherical symmetry.  The solid lines denote the numerical results of Section \ref{sec:SS}; the dashed lines represent the asymptotic scalings (\ref{beta_ana}) and (\ref{K_ana}).}
\label{Fig3}
\end{figure}

Under the assumption of spherical symmetry we may assume $\Omega_j = 0$ in (\ref{beta_ansatz}), even though we will find below that we can always relax this assumption.   We may also assume that the term $B^i$ in (\ref{beta_ansatz}) is purely radial.  In the following we abbreviate $\beta = \beta^r$ and $B = B^r$. Inserting this into (\ref{K}) yields
\begin{equation} \label{K_SS}
K \sim \frac{B}{r^m} \frac{M_\alpha}{r^2} = \frac{B M_\alpha}{r^{m+2}},
\end{equation}
where we have assumed $n \geq 1$ for the moment, so that the leading term for $\alpha$ is given by the monopole term.  To determine $m$ we have to consider the momentum constraint (\ref{betaConstraint}).  In spherical symmetry, this equation may be written as
\begin{equation}\label{betaConstraintMod}
\partial_r^2 \beta +  \left( \displaystyle \frac{2}{r} - \frac 
{\partial_r \alpha}{\alpha} + 6 \frac{\partial_r \psi}{\psi} \right)  
\left(\partial_r \beta  - \displaystyle \frac{\beta}{r}\right)   
= \alpha \partial_r K 
\end{equation}
(see equation (12b) in \cite{MatBG08}).  Restricting attention to the leading order terms this reduces to
\begin{equation}
\partial_r^2 \beta + \frac{2}{r}\left( \partial_r \beta - \frac{\beta}{r} \right) \sim 0.
\end{equation}
We can now insert the ansatz (\ref{beta_ansatz}) and find $m = 2$ as an asymptotically flat solution.  Inserting this into (\ref{K_SS}) shows that $K \sim r^{-4}$.   We graphically verify that our numerical solutions of Section \ref{sec:SS} satisfy this asymptotic fall-off behavior in Figure \ref{Fig3}.  From (\ref{matrixA}) we also see that $\bar A^{ij} \sim r^{-3}$ (which is consistent with the observation that in spherical symmetry, traceless-transverse tensors are unique up to an overall scaling factor, fall off with $1/r^3$, and can be computed from $(\bar L W)^i$ where $W^i$ is proportional to $(1/r^2, 0 , 0)$).

Finally, we can determine $n$ by inserting the decay rates of  $K$, $K^{ij}$ and $\beta$ into (\ref{eq:lapse}).  The dominant term is the $K^{ij}K_{ij}$ term which falls off like $1/r^6$. This yields $n=4$, which shows that our above assumption $n\geq1$ was self-consistent for our purposes here. 

We could have obtained the scaling for $\alpha$ alternatively by observing that  
\begin{equation} \label{prop_dist}
\alpha^2 - \beta^2 = 1 - \frac{2M}{R},
\end{equation}
where, as before, $R$ is the radial radius.  If $\alpha$ approaches unity at infinity, and $\beta \sim {\mathcal O}\left(r^{-2}\right)$, we must have
\begin{equation}
\alpha^2 = 1 - \frac{2M}{R} + {\mathcal O}\left(\frac{1}{R^4}\right)
\end{equation}
or
\begin{equation}
\alpha = \sqrt{1 - \frac{2M}{R}} + {\mathcal O}\left(\frac{1}{R^4}\right) 
\end{equation}
If the spatial metric is in the standard Schwarzschild form, it is a straightforward calculation to show that $D^2 (1 - 2M/R)^{1/2} = 0$. Therefore we have $D^2 \alpha = {\mathcal O}\left(r^{-6}\right)$, exactly as required.
 
In fact, we can also derive this asymptotic behavior analytically.
Combining equations (\ref{prop_dist}) and (\ref{first_int}) we find
\begin{equation} \label{beta_ana}
\beta \sim \frac{C e^{1/2}}{R^2} \approx \frac{2.055 M^2}{R^2}.
\end{equation}
We can now insert this into (\ref{1+log(2)}) to find
\begin{equation} \label{K_ana}
K \sim \frac{C e^{1/2}}{2} \frac{M^2}{R^4} \approx \frac{1.028 M^4}{R^4}.
\end{equation}
We include (\ref{beta_ana}) and (\ref{K_ana}) in Figure \ref{Fig3}, and find excellent agreement with our numerical results.

Before closing this Section we point out that there was no need to assume $\Omega_j = 0$ in (\ref{beta_ansatz}).  The corresponding rotation term in the shift is purely angular, but since all gradients must be purely radial in spherical symmetry, this term drops out both in (\ref{K}) and (\ref{eq:lapse}).   For example, if we orient the $z$-axis along $\Omega_i$, then the rotational term in the shift is $\beta^\phi$, but all derivatives along $\partial_\phi$ vanish in spherical symmetry.  This means that we could, if we wanted, allow for a rotating coordinate system when constructing stationary 1+log slices of a spherically symmetric spacetime.  This is a special case of a more general result for stationary axisymmetric systems which we will derive in Section \ref{sec:R+T}.   For more general rotating systems, however, for which only a helical Killing vector exists, this does not seem to be possible. This is the content of the next subsection.

\subsection{Rotation}

We first observe that spacetimes describing rotating systems that are not axisymmetric generally do not possess a timelike Killing vector (e.g.~they are radiating).  Therefore we should generally not be able to find an exact stationary slicing of such spacetimes.   Certain spacetimes of interest, e.g.~binary black holes in approximately circular orbit, nevertheless possess an approximate helical Killing vector.  For binaries, this helical Killing vector generates the orbit of the binary companions.  If such an approximate helical Killing vector exists, we could apply the above formalism to construct approximately stationary slicings of such spacetimes.   In Sections \ref{sec:1+log} and \ref{sec:CTS} we identified the lapse and shift with the Killing lapse and shift, and based on these choices we set the time derivatives of the conformally related metric and the mean curvature to zero.  This means that we assume the Killing lapse be of the form (\ref{alpha_ansatz}) and the Killing shift of the form (\ref{beta_ansatz}) with $\Omega_j \ne 0$.  In practice, this angular velocity is determined as part of the iterative algorithm that is used to construct the solution.

We again assume that we have found a solution, in which case the lapse $\alpha$ must satisfy equation (\ref{eq:lapse}).   As before we assume that all the quantities belong to some weighted spaces, i.e., that each derivative introduces an extra power of $r$ in falloff.  It is easy to see that, at least generically, equations (\ref{1+log(2)}) and (\ref{eq:lapse}) are incompatible at large $r$.  A simple counting argument goes as follows.  Let us assume that the right hand side of (\ref{eq:lapse}) falls off like $1/r^A$, say.  Then (\ref{eq:lapse}) implies that $\alpha$ must contain a term that decays like $1/r^{A-2}$ (the complementary part of $\alpha$ may contain terms that fall off even slower).   Now insert this information in (\ref{1+log(2)}). The term $\partial_i \alpha$ decays like $1/r^{A - 1}$, while $\beta$, of course, diverges as $r$. Therefore $K$ must decay like $1/r^{A-2}$.   When this is inserted back into (\ref{eq:lapse}), we see that $\beta^i \partial_i K $ only decays like $1/r^{A-2}$, thus contradicting our original assumption.\footnote{One might be concerned that the $\alpha K_{ij}K^{ij}$ and the $\beta^i \partial_i K$ terms in (\ref{eq:lapse}) could cancel each other, thereby avoiding the contradiction.  We can show that this is impossible with the help of the following argument.  Asymptotically, the shift $\beta^i$ is dominated by $\beta^i \rightarrow \epsilon^{ijk} \Omega_j x_k \equiv \beta^i_{\rm rot}$; let us therefore assume that $\alpha K_{ij}K^{ij}$ is canceled by $\beta^i_{\rm rot} \partial_i K$.  But $\beta^i_{\rm rot}$ is proportional to the $\ell = 1$, $m = 0$ {\em magnetic} vector spherical harmonic $S_a^{10}$, which has axial parity.  The angular components of $\partial_i K$ have polar parity, and can be expanded into {\em electric} vector spherical harmonics $E_a^{\ell m}$.  An integral of $\beta^i_{\rm rot} \partial_i K$ over any sphere must therefore vanish, indicating that $\beta^i_{\rm rot} \partial_i K$ itself must take both positive and negative values (unless it is identically zero).  The term $\alpha K_{ij}K^{ij}$, however, is non-negative, meaning that it cannot be canceled by the $\beta^i_{\rm rot} \partial_i K$ term everywhere.}

It is easy to see why the above argument does not hold for axisymmetric spacetimes.  Let us assume, to leading order, that $\beta$ is a rotation about the symmetry axis, which we align with the $z$-axis so that $\beta^i \sim (0, 0, \beta^\phi)$ to leading order.  But $\partial_\phi \alpha = 0 = \partial_\phi K$, since $\partial_\phi$ is a Killing vector.  Therefore $\beta^i\partial_i \alpha$ and $\beta^i \partial_i K$ vanish identically, rather than giving terms which decay slowly.  In the presence of a true rotational Killing vector, therefore, equations (\ref{1+log(2)}) and (\ref{eq:lapse}) are not incompatible, and it is possible to find asymptotically flat, stationary 1+log slices.   This is not surprising, of course, since dynamical simulations of binary black hole mergers, using the 1+log slicing, settle down at late times to a time-independent and axisymmetric solution describing a single Kerr black hole, which clearly must satisfy both (\ref{1+log(2)}) and (\ref{eq:lapse}).
  
To account for the special case of axisymmetry we can revise the above argument as follows.
Consider the nonaxisymmetric part of the right hand side of equation (\ref{eq:lapse}) only. It is this part which we assume decays like $1/r^A$. This implies that $\alpha$ has a non-axisymmetric part which decays like $1/r^{A-2}$. This, when substituted into (\ref{1+log(2)}), implies that $K$ has a nonaxisymmetric part which only decays like $1/r^{A-2}$. In turn, this means that the right hand side of equation (\ref{eq:lapse}) can only decay like $1/r^{A-2}$, in contradiction to our above assumption.
  
\subsection{Stationary and axisymmetric spacetimes}
\label{sec:R+T}

We now assume that the asymptotic behaviour of the spacetime is like that of the Kerr solution.  In particular we assume that we have two linearly independent Killing vectors, namely a time translation Killing vector $T^{a}_K$ and a rotational Killing vector $\Phi^{a}_K$.  We will further assume that a slice $\Sigma$ exists which is a stationary 1+log slice with respect to the timelike Killing vector $T_K^a$, and that this slice is axisymmetric in the sense that the rotational Killing vector lies in the slice.  We would then like to show that this slice is also a stationary 1+log slice with respect to any helical Killing vector 
\begin{equation} \label{H_K}
H^{a}_K = T^{a}_K + \Omega \Phi^{a}_K
\end{equation} 
where $\Omega$ is now an arbitrary constant.

We start by writing the Killing vectors in terms of their corresponding Killing lapses and shifts as 
\begin{equation} \label{T_K}
T^a_K = \alpha_T n^a + \beta^a_T
\end{equation} 
and 
\begin{equation} \label{R_K}
\Phi^a_K = \alpha_\Phi n^a + \beta^a_\Phi,
\end{equation}
where $n^a$ is the normal on $\Sigma$.  Since $\Phi_K^a$ is tangent to $\Sigma$, we must have 
$\alpha_\Phi = 0$.  Inserting (\ref{T_K}) and (\ref{R_K}) into (\ref{H_K}) we can therefore write the helical Killing vector as
\begin{equation}
H^a_K = \alpha_H n^a + \beta^a_H = \alpha_T n^a + \beta_T^a + \Omega \beta_\Phi^a
\end{equation}
and identify $\alpha_H = \alpha_T$ and $\beta_H^i = \beta_T^i + \Omega\beta_\Phi^i$.

Given that $\Sigma$ is a stationary 1+log slice with respect to $T_K^a$ we have
\begin{equation} 
K = \frac{\beta_T^i \partial_i \alpha_T}{2 \alpha_T} \label{T}.
\end{equation}
In order to show that $\Sigma$ is also stationary 1+log with respect to $H_K^a$ we evaluate
\begin{equation} 
K = \displaystyle \frac{\beta_H^i \partial_i \alpha_H}{2 \alpha_H} 
 = \frac{(\beta_T^i + \Omega \beta_\Phi^i) \partial_i \alpha_T}{2 \alpha_T}  
 = \displaystyle \frac{\beta_T^i \partial_i \alpha_T}{2 \alpha_T}  + 
 \frac{\Omega}{2 \alpha_T} \Phi_K^a \partial_a \alpha_T.  \label{H} 
 \end{equation} 
Now consider a coordinate system with two unit vectors $(e_T)^a$ and $(e_\Phi)^a$ that are aligned with the Killing vectors $T_K^a$ and $\Phi_K^a$.   In this coordinate system, the metric must then be independent of the coordinates $T$ and $\Phi$ associated with $(e_T)^a$ and $(e_\Phi)^a$.  Moreover, we have $\alpha_T = (-g^{TT})^{-1/2}$.   This implies that $\alpha_T$ must be independent of $\Phi$, and hence $\Phi_K^a \partial_a \alpha_T = 0$.  The last term in equation (\ref{H}) therefore vanishes, so that the condition (\ref{H}) holds if (\ref{T}) does.  This completes the proof (see also \cite{OM08}).

The above argument demonstrates both more formally and more generally an assertain that we made at the end of Section \ref{sec:bin-SS}, namely that there was no need to assume $\Omega = 0$ in our analysis of spherically symmetric spacetimes.  

\section{Summary and Discussion}
\label{sec:summary}

In this paper we propose and explore a ``stationary 1+log" slicing condition, defined by equation (\ref{1+log(2)}), for the construction of initial data.  For stationary spacetimes, this condition automatically yields a stationary foliation when the data are evolved with the gauge conditions used in moving puncture simulations.  The resulting slicing is time-independent and agrees with the slicing that is generated by dragging the initial data along the spacetime's time Killing vector.  This is a desirable property, since it avoids pure ``coordinate" evolutions that may otherwise introduce numerical error and noise, and that may contribute to the ``spurious gravitational radiation" that is often observed at early times in binary black hole simulations.  Given these considerations, this slicing condition would appear as an attractive alternative to maximal slicing, which is often used in the construction of binary black hole initial data.

Unfortunately, it does not seem to be possible to construct initial data for binary black holes -- or any binaries -- in the stationary 1+log slicing.  Spacetimes containing binaries in approximately circular orbit possess an approximate helical Killing vector that generates the binary's orbit.   We provide an argument that shows that 1+log slices that are stationary with respect to such a helical Killing vector cannot be asymptotically flat, unless the spacetime possesses an additional axial Killing vector.   As a consequence, the stationary 1+log slicing condition, without modification, seems useful only in the context of axisymmetry spacetimes, for example for rotating neutron stars.

Our results also apply to generalizations of the 1+log slicing (\ref{1+log_dyn}), for which the right hand side $- 2 \alpha K$ is replaced by $- n f(\alpha) K$, where $n$ is some arbitrary number and $f(\alpha)$ some non-zero and finite function of $\alpha$ (see \cite{BonMSS95}).  

It may be possible, however, to modify the stationary 1+log slicing condition in such a way that it can be applied to binaries as well.  In particular, it seems promising to apply (\ref{1+log(2)}) only locally in a neighborhood of the binary, and match to maximal slicing $K=0$ asymptotically.  Such an approach would still benefit from the slicing being stationary in the neighborhood of the binary, where the gravitational fields are strongest, but would avoid the problems associated with the asymptotic behavior of stationary 1+log slices.  We plan to pursue this approach in the near future.

\ack

We would like to thank Mark Hannam for many useful discussions.
KM gratefully acknowledges support from the Coles Undergraduate  
Research Fellowship Fund at Bowdoin College and from the Maine Space  
Grant Consortium.   This work was supported in part by NSF grant  
PHY-0456917 to Bowdoin College, SFI grant 07/RFP/PHYF148 to NOM,  
NASA grants NNG04GK54G and NNX07AG96G to UIUC, and NSF grants PHY-0650377,
PHY-0345151 and PHY0205155 to UIUC.
 
\appendix 
 
\section{Dynamical evolution of Schwarzschild}
\label{sec:SS_evolve}

 In this Appendix we present some results for dynamical evolution calculations for Schwarzschild spacetimes, starting both with ``stationary 1+log" initial data (``trumpet data") and maximally sliced puncture data (``wormhole data").  While these results are not of direct relevance for the main arguments laid out in the paper, and while, being restricted to spherical symmetry, their scope is limited, they do demonstrate the potential of the stationary 1+log slicing condition to reduce numerical error.

Dynamically evolving the initial data of Section \ref{sec:SS} with the moving puncture method adopted in our code leads to one complication that is related to the black hole interior.  Using the conformal thin-sandwich formalism for the construction of the initial data we have excised the interior by imposing boundary conditions on the black hole horizons.  The initial data therefore cover only the black hole exterior.  The moving puncture method, on the other hand, requires data everywhere, including the interior.  One way to solve this problem is to fill the black hole interior with some arbitrary choices for the metric functions.  As long as the constraint-violating data in the interior do not propagate to the exterior, this ``junk'' will not affect the outside of the black hole \cite{EtiFLSB07,BroSSTDHP07}, at least in the continuum limit.  In experiments for Schwarzschild black holes with the gauge conditions used in moving puncture evolutions, even the interior settles down to a constraint-satisfying solution that is determined by the exterior data, but not the initial choices inside the black hole.

One source of error in this method arises from the junction of the exterior data to the arbitrary ``junk" inside.  In our finite-difference code, a grid stencil that is centered on a grid-point just outside the horizon may extend into the black hole interior.  Numerically evaluated derivatives are therefore sensitive at these grid-points to the particular choices of the interior ``junk", and to how it is matched to the exterior solution (see \cite{EtiFLSB07,BroSSTDHP07}).   The juncture of these data therefore results in a small numerical error, directly originating from the black hole horizon, that can be controlled by improving the continuity of the data.  In our implementation we construct the interior  ``junk" data using an 8th-order polynomial extrapolation of the exterior data into the interior.

\begin{figure}[t]
\includegraphics[width=3in]{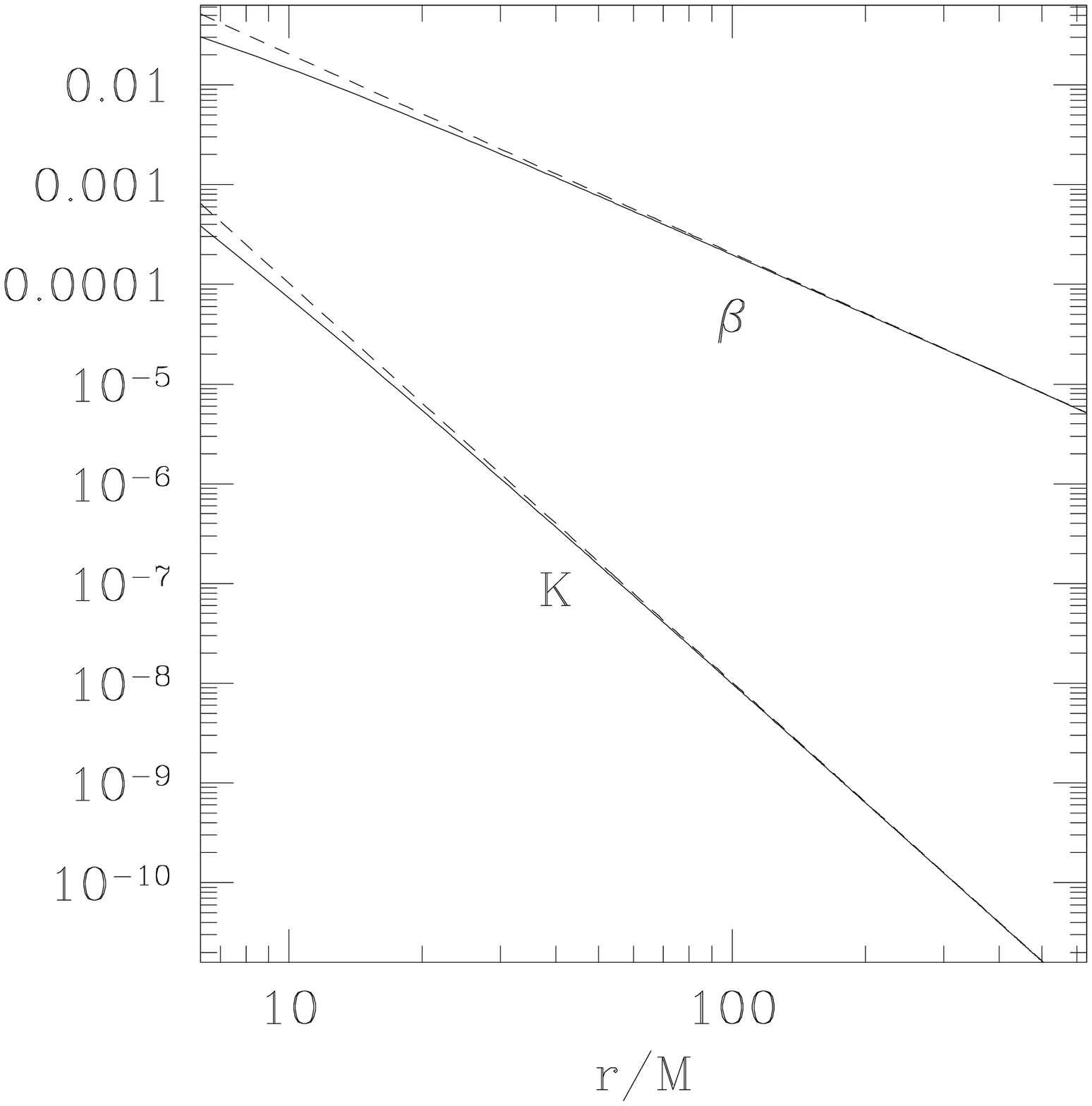}
\caption{Relative changes in the ADM mass, $\Delta M = (M(t) - M_0)/M_0$, where $M_0$ is the initial ADM mass at $t=0$, and where we compute the ADM mass by performing a surface integral on a sphere of coordinate radius $r = 45.6M$.  The solid line is the result of an evolution starting with puncture initial data representing a slice of constant Schwarzschild time (``wormhole" data), while the dashed line results from the evolution of our stationary 1+log initial data (``trumpet" data).}
\label{Fig2}
\end{figure}

In Fig.~\ref{Fig2} we compare the dynamical evolution of the normal puncture initial data, i.e.~a slice of constant Schwarzschild time expressed in isotropic coordinates (``wormhole" data), with that of our data in stationary 1+log slicing (``trumpet" data).  In both cases we use our three-dimensional code (i.e.~we purposely do not take advantage of the spherical symmetry) that has been described in detail in \cite{FabBEST07,EtiFLSTB08}.   As gauge conditions we use the 1+log slicing (\ref{1+log_dyn}) together with the ``unshifting shift" version of the ``$\bar \Gamma^i$-freezing" condition 
\numparts
\begin{eqnarray} \label{eq:gamma_driver}
\partial_t \beta^i & = & (3/4) B^i \\
\partial_t B^i & = & \partial_t \bar \Gamma^i - \eta B^i
\end{eqnarray}
\endnumparts
with $\eta = 1.0/M$, where $M$ is the black hole's ADM
mass and ${\bar \Gamma^i} = -\partial_j {\bar \gamma}^{ij}$.
As initial conditions for these gauge conditions we choose $B^i = 0$ as well as the Killing lapse and shift, as found from the solution of the conformal thin-sandwich equations.  If $t^a$ is aligned with a Killing vector of the spacetime, then $\beta^ i =$ const is a solution to the shift condition (\ref{eq:gamma_driver}).

Our code uses fourth-order spatial differencing and a fourth-order Runge-Kutta 
time stepping via the method of lines.  We use the {\tt Carpet} 
infrastructure~\cite{carpet} to implement the fixed mesh refinement (FMR) with nine levels of refinement.  We resolve the black holes at the finest level with a resolution of $M/32$, and our grid extends to $256M$ at the outer boundary.  

As a measure of the numerical error we show in Fig.~\ref{Fig2} relative changes in the ADM mass, evaluated as the surface integral 
\begin{equation}
M = -  \frac{1}{2\pi} \int_{S} \left(\bar D^i \psi - \frac{1}{8} \bar \Gamma^i
\right) d  \bar S_i
\end{equation}
on a sphere $S$ of radius $r = 45.6M$.  Specifically, we graph $\Delta M = (M(t) - M_0)/M_0$, where
$M_0$ is the initial ADM mass at $t=0$.  Strictly speaking, the ADM mass is defined only at spatial infinity.  In numerical relativity codes whose grids do not extend to spatial infinity it is a common practice to evaluate an approximate value of the ADM mass at a finite separation instead.  The resulting approximations converge to the correct ADM mass as the extraction radius approaches infinity; assuming reasonable fall-off properties this can be seen easily by using the Hamiltonian constraint to write the ADM as a volume integral.   The errors shown in Fig.~\ref{Fig2} are therefore, strictly speaking, not errors in the ADM mass, but close approximations and indicative of what would be observed in a dynamical simulation.

Two observations are noteworthy.  Firstly, adopting initial data on a slice of constant Schwarzschild time leads to deviations of the numerically determined ADM mass from its correct value that are larger than those resulting from the stationary 1+log data.  These errors are a consequence of the ``wormhole" data collapsing into ``trumpet" data.  Secondly, we notice that these errors arrive at the extraction radius $r = 45.6 M$ earlier than those for the stationary 1+log data.   The largest source of error for the ``trumpet" data are those arising on the black hole horizon, as discussed above.  As one might expect, these errors arrive at the extraction surface after a light-travel time of approximately $t = 45M$.  For the ``wormhole" data, on the other hand, the errors arrive earlier, because the transition from ``wormhole" data to ``trumpet" data occurs everywhere, including the exterior of the black hole (even though the size of these transitions decreases rapidly with the distance from the black hole).  

We also monitored the black hole's irreducible mass.  While the relative deviations from the exact value were very small in both simulations (namely on the order of $10^{-5}$), the errors resulting from the ``wormhole" initial data were about twice as large as those resulting from ``trumpet" initial data.

\end{document}